\documentstyle[epsf,referee]{l-aa}
\newlength{\fsize}
\begin{document}

\thesaurus{12 (12.03.3; 12.07.1; 13.18.1; 11.12.2)} 

\title{Cosmological Parameters from Statistics of Strongly Lensed Radio Sources}

\author{Asantha R. Cooray}
\institute{Department of Astronomy and Astrophysics, University of Chicago, Chicago IL 60637, USA. \\ E-mail: asante@hyde.uchicago.edu}

\date{Received: October 26, 1998  ; accepted: November 27, 1998 }
\maketitle

\begin{abstract}
We calculate the expected number of strongly lensed radio sources
in a sample of $\sim$ 6500 sources
observed with the Very Large Array 
as part of the Cosmic Lens All Sky Survey  (CLASS)
during the first two sessions of its observations.
A comparison between the predicted and
the observed number of lensed radio sources allows a 
determination of the current value of 
$\Omega_m-\Omega_\Lambda$, where $\Omega_m$ is
the cosmological mass density of the universe and $\Omega_\Lambda$ is
the normalized cosmological constant. 
If there are six strongly lensed sources in this sample,
our 95\% confidence lower
limit on $\Omega_m-\Omega_\Lambda$ is -0.58.
For a flat universe with $\Omega_m+\Omega_\Lambda=1$, 
then, $\Omega_\Lambda < 0.79$ (95\% C.L.). If there
are ten strongly lensed sources, the 95\% confidence lower limit
on $\Omega_m-\Omega_\Lambda$ is -0.90.
These lower limits are consistent with estimates
based on high redshift supernovae and with previous
limits based on gravitational lensing.
Instead of considering a simple cosmological constant, we also consider the
possibility of a quintessence scalar field responsible for the
additional energy density of the universe, with an equation of
state of the form $w=P_x/\rho_x$, where
$P_x$ and $\rho_x$ are the pressure and energy density of the field. 
We present our constraints on the
$\Omega_x-w$ plane, where $\Omega_x$ is the present
day normalized energy density of the scalar-field component, 
assuming a flat universe such that $\Omega_m+\Omega_x=1$. 
If there are 6 strongly
lensed sources in the present CLASS sample, 
gravitational lensing statistics allow us to rule
out the region with $\Omega_x \ga (1.2 - 0.5 w^2) \pm 0.05$ (95\% C.L.).
We discuss the region allowed by combined
gravitational lensing statistics,  high redshift Type Ia supernovae distances,
and globular cluster ages.

Instead of a cosmological model, we can  constrain the redshift
distribution of faint radio sources based on the observed
gravitational lensing rate and an assumed cosmological model. 
If there are six strongly lensed sources, the 68\% confidence
upper limit on the average redshift $\langle z \rangle$
of radio sources with flux
densities less than 150 mJy at 8.4 GHz is 
$ \langle z \rangle < 1.4 + (\Omega_m-\Omega_\Lambda) \pm 0.1$.
In order to obtain a much tighter estimate on the cosmological parameters,
it is essential that the redshift distribution for radio sources at the 
faint flux density levels be observationally 
determined. We strongly recommend that statistically complete 
optical spectroscopic programs be carried out to
obtain  redshifts
for a representative subsample of faint background radio sources.
Until such redshifts are obtained, it is unlikely that a major
improvement could be made with respect to lensed radio source constraints
on cosmological parameters.

\end{abstract}
\section{Introduction}

The number of strongly lensed sources found in individual
observational programs can be used to constrain the cosmological
constant (e.g., Turner 1990). Using statistics of optical
lensing programs, various  studies have now derived
upper limuts on the present day value of the
 cosmological constant (Kochanek 1996a;
Chiba \& Yoshi 1997; Cooray et al. 1998a). As discussed
in the literature, however, lensed source search programs 
at optical wavelengths can be affected by
systematics arising from effects due to extinction and 
reddening  (e.g., Malhotra et al. 1996), and
issues related to completeness (Kochanek 1991). 
Such effects, especially extinction for which there is no 
 clear consensus, cannot easily be quantified in 
cosmological studies with gravitational lensing statistics.
At radio wavelengths, neither of these effects are 
likely to affect the statistics (see, Falco et al. 1998),
including completeness,  when flux-limited systematic
surveys are carried out.
Also, high resolution interferometers such as the VLA and  MERLIN allow
detection of lensed sources with image separations much smaller than
that allowed by ground-based optical telescopes.
Thus, statistics from lensed radio source surveys
 are generally preferred over the optical ones
to obtain cosmological parameters.

Recently, Falco et al. (1998) studied the statistics of lensed radio sources
in the Jodrell Bank-VLA Astrometric Survey (JVAS; Browne et al.
1997; Patnaik et al. 1992) sample of
2500 flat-spectrum sources brighter than 200 mJy at 5 GHz.
They were able to derive a 2$\sigma$ lower-limit on the mass density of
the universe of $\Omega_m > 0.27$ ($0.47 < \Omega_m < 1.38$ at 1$\sigma$).
Recently, the number of strongly lensed radio sources has steadily increased
 with results from the  Cosmic Lens All Sky Survey (CLASS), which has  
observed $\sim$ 6500 flat-spectrum sources during the first two sessions of
 its observations using the VLA.
The total number of currently confirmed lensed radio sources 
from these observations amount to six: B1608+656 
(Myers et al. 1995), B2045+265 (Fassnacht et al. 1998), B1933+503 (Sykes 
et al. 1998), 
B0712+472 (Jackson et al. 1998a), B1600+434 (Jackson et al. 1995),
B1127+385 (Koopmans et al. 1998). The CLASS  is a collaboration between groups
at Jodrell Bank, Caltech, Dwingeloo and Leiden, and is an extension  to
the previous JVAS by observing
flat-spectrum sources ($\alpha \geq -0.5;
S_\nu \propto \nu^\alpha$) down to a flux density limit of 30 mJy
at 5 GHz.

Here, we calculate the expected number of lensed radio sources in the
CLASS sample of $\sim$ 6500 sources
 for different cosmological parameters, and constrain
these parameters by comparing the predictions with the observations.  
In Sect. 2, we discuss our calculation and its inputs,  including
the redshift distribution for background
radio sources and the luminosity  distribution
of foreground lenses.  In Sect. 3, we present our resulting constraints on
cosmological parameters, and in particular on
$\Omega_m-\Omega_\Lambda$, as well as
on the redshift distribution of the faint radio sources.  
In Sect. 4, we discuss the potential effect of
systematic errors in the predicted and observed rate of strong
lensing.  Finally, in Sect. 5, we summarize and discuss future prospects for
tighter constraints. We follow the conventions that the Hubble
constant, $H_0$, is 100\,$h$\ km~s$^{-1}$~Mpc$^{-1}$, the present normalized 
mean density of the universe $\Omega_{m}$ is $8 \pi G \rho_m/3 H_0^2$,
and the present normalized cosmological constant $\Omega_\Lambda$ is 
$\Lambda/3 H_0^2$.  
In a flat universe, $\Omega_m+\Omega_\Lambda=1$. We have simplified the 
notation by not including the subscript '0'
usually taken to denote the present day values of these quantities.

\section{Expected number of strongly lensed sources}

\subsection{Calculational method}

Our present calculational method is quite similar  to that of Cooray
et al.
(1998a; hereafter CQM), in which
we calculated the expected lensing rate in the Hubble Deep Field (HDF;
Williams et al. 1996),
as implied by various photometric redshift catalogs for HDF sources.
A comparison between the observed number of lensed sources in the HDF and
the predicted number, allowed us to
calculate the present value of $\Omega_m-\Omega_\Lambda$.
We follow a procedure similar to CQM by first establishing the
redshift distribution of the observed CLASS sources, and then by
using the standard lensing  
calculations to estimate the expected number of strongly
lensed radio sources present in this sample.

In Fig.~1, we show the flux density 
distribution of CLASS sources. Except for a 
handful of sources with flux densities greater than 200 mJy, 
most of the observed sources follow a 
 distribution that sharply rises around 30 mJy. This is
only an observational artifact due to the 30 mJy cutoff imposed on
the initial source selection. Since the source sample was selected based on
previous Green Bank 5 GHz surveys  (e.g., Gregory et al. 1996)
and that we are now using the measured flux densities
from CLASS VLA snapshot 
maps at 8.4 GHz, the presence of sources with flux densities
less than 30 mJy at 8.4 GHz suggest that the current CLASS sample includes
a small number of non-flat spectrum radio sources; The CLASS definition
with respect to flat spectrum sources is based on an upper frequency point
at 5 GHz rather than 8.4 GHz. Sources with flux densities
less than 30 mJy are also expected
given uncertainties in the Green Bank survey flux density measurements
 and the possibility for intrinsic flux density variation
between Green Bank and CLASS survey epochs. 
The total number of radio sources in the publicly available catalog
of CLASS sources\footnote{http://www.jb.man.ac.uk/$\sim$njj/glens/class.html} during the first two sessions is 6528.

\begin{figure}[ht]
\vspace{6cm}
\caption{The flux density 
distribution for CLASS sources. We have only shown the
distribution of sources with flux densities less than 150 mJy, where
most of the sources are found}
\end{figure}                                                            

As discussed in Falco et al. (1998), one of the main problems associated
with lensed radio source statistics is the non-availability of redshifts 
for radio sources. However, it has also 
been shown by Falco et al. (1998), and previously by Kochanek (1996b), 
that the radio sources exhibit a strong correlation between their redshift
and flux density distributions.
Thus, even though complete redshift surveys may not exist, 
it is still possible to construct with existing data 
a statistically accurate redshift distribution for a given radio flux density. 
This assumes that the correlation between redshift and flux
density distribution remains to be valid at flux levels
lower than observationally studied.
Since redshifts cannot be determined exactly for individual
sources, each of the
assigned redshifts at a given flux density will have an uncertainty
which can be accounted when
calculating cosmological parameters. Also, since the CLASS radio source
sample is substantially large, on average, the modeled redshift distribution 
can be expected to represent the true redshift 
distribution. Since no new radio source redshift surveys 
have been published recently,
we use the parameterization of the flux$-$redshift relation 
in  Falco et al. (1998)
to calculate the probable redshifts
of CLASS sources. For each redshift, we also assign
an uncertainty based on the error in the parameterization
presented by Falco et al. (1998; see, their Fig.~ 4).
Since no redshift surveys exist for radio sources with flux densities
less than  $\sim$ 200 mJy, we use the extrapolation of the flux$-$redshift
distribution to estimate the redshifts of these sources.
However, we assign
a slightly higher uncertainty for these redshifts than expected from the
simple extrapolation presented in Falco et al. (1998). 
We assume that the model A in Falco et al. (1998) is the mean
redshift distribution, while models B and C, with an increase in
uncertainty, represents our higher and lower redshift distributions,
respectively. With redshift and flux density for each source, we calculate
the radio luminosity, $L$,  and its uncertainty. We do not consider any errors
in the flux density measurement as these 
are likely to be small compared to the ones associated with
the lensing calculation.

In order to calculate the lensing rate for radio sources, we model the
lensing galaxies as singular isothermal spheres (SIS) and use the
analytical  filled-beam approximation (e.g., Fukugita et al.
1992). At redshifts $z \leq 4$, the analytical filled-beam
calculation  in Fukugita et al. (1992) agree to better than 2\% with
numerical calculations (e.g.,  Fig.\ 1 in Holz et al. 1998).

Gravitational lensing statistics are 
affected by the so-called ``magnification bias'' (Kochanek 1991) 
in which the number of lensed sources in the
sample can be different from an unlensed sample down to the same flux density
level. Thus, any calculation involving
lensed source statistics should account for the magnification bias
and associated systematic effects. 
The effect of this bias, whether to  increase or decrease the number
of lensed sources, depends
on the slope of the number counts.
The magnification bias is
particularly pronounced in quasar lensing
surveys (Maoz \& Rix 1993), because the faint end of the
quasar luminosity function rises steeply, increasing the number
of lensed sources.   In Sect. 2.2, we discuss this effect
for radio sources based on the flat-spectrum radio luminosity
function.
                               
Following CQM, if the probability for a source at redshift $z$ to
be strongly lensed
is $p(z,\Omega_m,\Omega_\Lambda)$, as calculated
based on the filled-beam equation, we can write the
number of lensed sources, $\bar N$,  as (see, also, Maoz et al. 1992):
\begin{eqnarray}
\bar N = & \sum_i p(z_i,\Omega_m,\Omega_\Lambda) B(L_i,z_i) g(\Delta \theta,\Delta f) \nonumber \\
 & \equiv \sum_i \tau(z_i) \;,                              
\end{eqnarray}
where $B(L_i,z_i)$ is the magnification bias for a radio source
at redshift $z$ with luminosity $L$, and $g(\Delta \theta,\Delta f)$ is 
the selection function
defined based on image separation, $\Delta \theta$, and flux 
ratio, $\Delta f$,
between components of the strongly lensed sources.
Here, the sum is over each of the radio sources in our sample.
The index $i$ represents each object; hence, $z_i$ and $L_i$ are,
respectively, the redshift, and the  rest-frame luminosity of the $i$th
source. 

For simplicity, following Kochanek (1996b), we assume that
the selection function is simply a uniform function in the range
$0\farcs3 \leq \Delta \theta \leq 6\farcs0$, such that all lensed
sources in this range are recovered.
For the SIS model, the fraction of lensed sources 
in this range of image separations is
0.917 (e.g., Kochanek 1996b). We have assumed that the image separation,
$\Delta \theta$, under the SIS assumption for foreground galaxies is
given by $\Delta \theta = 8 \pi (\sigma/c)^2$, where
$\sigma$ is the velocity dispersion of the foreground lensing
galaxies, and that the characteristic velocity dispersion for
a $L^*$ galaxy, $\sigma^*$, is 220 km s$^{-1}$ (see, Eq.~22 in Kochanek 1996b). 
Given that the observations are carried out to a much lower flux 
density level than the flux density of the observed source,  and that the
candidate lensed sources are selected when flux  ratios between strong
components are $\leq$ 10 (Browne et al. 1997),  it is likely that the
selection function is independent of the flux  ratio between
lensed images. For SIS models, the fraction of lensed sources with
image flux ratios $\leq$ 10 
is 0.9997,  and thus, for the purpose of this calculation we take 
$g(\Delta \theta,\Delta f)$ to be a constant with numerical value of 0.916.

\subsection{Magnification bias for radio sources}

The bias, $B(L,z)$, can be calculated based on the radio
luminosity function for background sources at redshift $z$, $\Phi(L,z)$:
\begin{equation}
B(L,z) = \frac{ \int_{A_{\rm min}}^{\infty} \Phi(L/A,z)\, P(A)\, dA\, d(L/A)}{\Phi(L,z)\, dL},
\end{equation}
where $P(A)dA$ is the probability distribution of amplifications $A$.
The GHz radio luminosity function for flat-spectrum sources, 
and its evolution,
is still not well determined observationally. 
For the purpose of 
the present paper, we follow the work of Dunlop \& Peacock (1990),
where luminosity
functions are calculated for sources at 2.7 GHz. Since we are 
dealing with a sample of flat spectrum sources, we can safely assume
that the 1 to 10 GHz luminosity function is equivalent to that
at 2.7 GHz.
We use the pure luminosity-evolution model (PLE) for flat-spectrum
sources, which is motivated by optical
quasar and X-ray AGN luminosity functions (e.g., Boyle et al. 1987):
\begin{equation}
\Phi(L,z) = \Phi_0 \left[ \left(\frac{L}{L_c(z)}\right)^{\alpha_1^*}+\left(\frac{L}{L_c(z)}\right)^{\beta_1^*}\right]^{-1},
\end{equation}
where $\alpha_1^*$ and $\beta_1^*$ are the
 faint and bright end slopes of the luminosity
function with respect to the critical, or break, luminosity $L_c$.
For the purpose of this paper,
the quantity $L$ is the radio luminosity (in W Hz$^{-1}$ sr$^{-1}$).
Since this luminosity function is defined in terms of an
element $d\log L$ instead of the $dL$ required in Eq.~2,
following King \& Browne (1996), we rewrite the
luminosity function in terms of a linear element, with new
slopes $\gamma_1$ and $\gamma_2$ below and above the critical luminosity:
\begin{equation}
\Phi(L,z) = \frac{\Phi_0}{L_c(z) \ln 10} \left[ \left(\frac{L}{L_c(z)}\right)^{\gamma_1}+\left(\frac{L}{L_c(z)}\right)^{\gamma_2}\right]^{-1}.
\end{equation}

The critical
luminosity, $L_c$, has been found to evolve with the form:
\begin{equation}
\log L_c(z) = a_0 + a_1 z + a_2 z^2.
\end{equation}
In Table~1, we list each of the coefficients.

\begin{table}
\caption[]{The adopted luminosity function for  radio sources
assuming a pure luminosity evolution model based on
Dunlop \& Peacock (1990)}
\begin{flushleft}
\begin{tabular}{cc}
\noalign{\smallskip}
\hline
\noalign{\smallskip}
Parameter & Value \\
\noalign{\smallskip}
\hline
\noalign{\smallskip}
Log($\Phi_0$/Mpc$^{-3}\Delta$ Log L$^{-1}$) & -8.15 \\
$\alpha_1^*$ & 0.83 \\
$\beta_1^*$  &  1.96 \\
$\gamma_1$  &  1.83 \\
$\gamma_2$  &  2.96 \\
$a_0$    &    25.26 \\
$a_1$    &   1.18 \\
$a_2$    &   -0.28 \\ 
\noalign{\smallskip}
\hline
\end{tabular}
\end{flushleft}
\end{table}

Since we are using the SIS model, the minimum
amplification, $A_{\rm min}$, is 2 and the probability distribution
for amplifications, $P(A)$, is $8A^{-3}$ (see, e.g., Maoz \& Rix 1993).
The magnification bias can  be written as (Maoz \& Rix 1993):
\begin{equation}                                                               
B(L,z) = \cases{8\left(\frac{L}{L_c}\right)^{\gamma_2-3}\left(\frac{1}{3-\gamma_1}-\frac{1}{3-\gamma_2}\right)+\frac{2^{\gamma_2}}{3-\gamma_2}, & $L > 2 L_c$; \cr
\frac{2^{\gamma_1} \left(\frac{L}{L_c}\right)^{(\gamma_2-\gamma_1)}}{3-\gamma_1}, & $L_c  < L < 2 L_c$; \cr
\frac{2^{\gamma_1}}{3-\gamma_1}, & $L_c  > L$. \cr }
\end{equation}

In Fig.~2, we show the magnification bias as a function of radio 
luminosity using 
the luminosity range for most of the sources in the CLASS Survey. For
sources  with luminosities such that $L(z) \leq L_c(z)$, 
the magnification bias is simply determined by the slope $\gamma_1$
and is numerically equivalent to $\sim$ 3.04.

\begin{figure}[ht]
\vspace{6cm}
\caption{The calculated $B(L,z)$ vs. $L$ for sources at
redshifts, $z$, of 0.5 (solid), 1 (dashed) and 1.5 (dotted) and 2
(dot-dashed). When $L(z) \leq L_c(z)$, 
the magnification bias is a constant determined by the slope of 
the lower end of the luminosity function $\gamma_1$}
\end{figure}

\subsection{Properties of lensing galaxies}

The probability of 
strong lensing depends on the number density and
typical mass of lensing galaxies.  For singular isothermal spheres,
this factor is conveniently represented by the dimensionless parameter 
(Turner, Ostriker \& Gott 1984):
\begin{equation} F\equiv
16\pi^3n_0R_0^3\left(\sigma\over{c}\right)^4\; .
\end{equation} Here $n_0$ is the number density of galaxies, $R_0\equiv
c/H_0$, and $\sigma$ is the velocity dispersion.  The
parameter $F$ is independent of the Hubble constant, because the
observationally inferred  number density is proportional to $h^3$.  We
can estimate $F$ at a given redshift from the galaxy luminosity
function, which we describe based on the Schechter function (Schechter 1976),
in
which the comoving density of galaxies at redshift $z$ and with
luminosity between $L$ and $L+dL$ is:
\begin{equation}
\phi(L,z)\, dL=\phi^*(z)\left[L\over{L^*(z)}\right]^{\alpha(z)}
e^{-L/L^*(z)}\, dL\;.
\end{equation}
In order to relate velocity dispersion,
$\sigma$, with luminosity, we assume a dependence between
absolute magnitude, $M$, and $\sigma$ of the form:
\begin{equation}
-M = a + b \log \sigma,
\end{equation}
which is known as the Faber-Jackson relation for early-type galaxies
(Faber \& Jackson 1976) and the Tully-Fisher relation for spiral galaxies (Tully \& Fisher 1977).

Thus, using Eqs.~8 and 9, Eq.~7 can be written as (Fukugita \& Turner 1991):
\begin{equation}
F= \frac{16 \pi^3}{c H_0^3} \phi^* \sigma_*^4 \Gamma\left(\alpha+\frac{10}{b}+1\right)\; ,
\end{equation}
where $\Gamma$ is the normal gamma function and $\sigma^*$ is given
by the characteristic magnitude $M^*$: $-M^* = a +b \log \sigma^*$.

To estimate $F$, we use a recent determination of
the foreground galaxy 
luminosity function by Zucca et al. (1997) based on 
observations of 3342 galaxies in the ESO Slice Project (ESP).
A comparison between the ESP luminosity function and 
the ones previously used in
lensing statistics calculations (e.g., Kochanek 1996, Falco et al. 1998) 
suggest no large variations. Thus, we expect our $F$ parameter
to be within the uncertainty of previous $F$ parameter estimations.
However, we note one change in our estimate for $F$ from a 
recent study of lensed source
statistics: We consider lensing galaxies to be composed of
both early-type as
well as spiral galaxies. This is contrary to the recent suggestion by
Chiba \& Yoshi (1998) that spiral and dwarf-type
galaxies do not contribute to the observed lensing rate.
We argue against this possibility simply based 
on the observational data, where
spiral galaxy lenses have been found
(e.g., B1600+434 in the present CLASS sample, Jaunsen \& Hjorth 1997; 
Koopmans et al. 1998). 
Zucca et al. (1997) find that  the 
luminosity function for galaxies is represented by a Schechter
function with parameters:
\begin{eqnarray}
M^*_B= & -19.61^{+0.06}_{-0.08}\; ,\qquad \alpha=-1.22^{+0.06}_{-0.07}\; ,\qquad  \phi^*=0.020 \pm 0.004 \ h^3{\rm Mpc}^{-3}\; .
\end{eqnarray}
These parameters agree with the recent estimation of the 
luminosity function of known lenses (Kochanek et al. 1998; see their Fig.~4).

In order to derive the characteristic velocity dispersion $\sigma^*$
for the observed $M^*$, we use the following Faber-Jackson and Tully-Fisher
relations:
\begin{eqnarray}
-M_B^\ast + 5\log h &=& (19.37 \pm 0.07) + 10 (\log \sigma_E^\ast - 2.3) 
   \qquad \mbox{\rm for E} \nonumber \\
-M_B^\ast + 5\log h &=& (19.75 \pm 0.07) + 10 (\log \sigma_{S0}^\ast - 2.3)
   \qquad \mbox{\rm for S0} \nonumber \\
-M_B^\ast + 5\log h &=& (19.18 \pm 0.10) + (6.56 \pm 0.48) (\log \sigma_{Sp}^\ast - 2.05) \qquad \mbox{\rm for Sp},
\end{eqnarray}
as presented by de~Vaucouleurs \& Olson (1982) for early-type galaxies (E/S0)
and Fukugita et al. (1991) for spiral galaxies (Sp).
To estimate the overall $F$ from individual contributions,
we use the galaxy fractions presented by Postman \& Geller (1984)
with ratios E:S0:Sp=12$\pm$2:19$\pm$4:69$\pm$4. These
ratios also agree with the relative numbers of spiral and 
early type galaxies in the Southern Sky Redshift Survey
(Marzke et al. 1998).
In Table~2, we list these parameters and estimate for $F$ using the
luminosity function from the ESP survey.

\begin{table}
\caption[]{$F$ value based on the ESP luminosity function}
\begin{flushleft}
\begin{tabular}{cccc}         
\noalign{\smallskip}
\hline
\noalign{\smallskip}
Type & Fraction & $\sigma^*$ & $F$ \\            
\noalign{\smallskip}
\hline
\noalign{\smallskip}
E  & 0.12 $\pm$ 0.02 & $210^{+10}_{-11}$ & 0.009 $\pm$ 0.004 \\
S0 & 0.19 $\pm$ 0.04 & $194^{+12}_{-10}$ & 0.010 $\pm$ 0.003 \\ 
Sp & 0.69 $\pm$ 0.04 & $131^{+15}_{-14}$ & 0.006 $\pm$ 0.004 \\
Total &             &                   & 0.026 $\pm$ 0.006 \\
\noalign{\smallskip}
\hline
\end{tabular}
\end{flushleft}
\end{table}

The best estimate for $F$ is 0.026 $\pm$ 0.006, with a relative
contribution of 31\% from spiral galaxies. Such a spiral
galaxy contribution agrees with the estimate by
Helbig (1998) that 2 of the 
6 known lensed CLASS sources are due to foreground spirals.
This observation also justifies our inclusion of
spiral galaxies in the $F$ parameter estimation.

Our estimate for $F$ is 
a factor of 2.5 higher than the value used by Chiba \& Yoshi (1998).
According to Kochanek et al. (1998), the two luminosity functions
used by Chiba \& Yoshi (1998) are the two models most
discrepant with the luminosity function from known lenses.
Using a low $F$ value can decrease the number of expected lensed sources
 enhancing the significance of
a cosmological constant in producing strongly lensed sources. 
In order to obtain reliable estimates of the cosmological parameters
based on statistics of lensed sources, as well as for comparative 
purposes, a consistent value of $F$ is needed among different studies. 
We note that our estimate for $F$ is consistent with previous 
estimates by Kochanek (1996a,1996b) and Falco et al. (1998).
Since there could be additional errors,
we allow for a slightly higher uncertainty (30\%)
and take $F$ to be 0.026 $\pm$ 0.008. This estimate
for $F$ is factor of four less than the value we recently estimated
for gravitational lensing in the HDF (CQM). This was both due to
the factor of 2 higher number density of sources in the HDF as suggested
by its luminosity function (Sawicki et al. 1997) and our partly incorrect
assumption that all such sources are elliptical galaxies.
The revised constraints on cosmological parameters from HDF lensing
rate based on a more appropriate $F$ value could
be found in Cooray et al. (1998b).

We calculate the expected number of  lensed sources by 
first estimating the redshift, luminosity and lensing rate
for individual CLASS sources
as a function of cosmology, with luminosity
used to account for the magnification bias.
We also vary the redshift according to flux$-$redshift correlation
and recalculate the expected number of strongly lensed sources.
In Fig.~3, we show the
expected number of strongly lensed radio sources in the CLASS survey
as a function of $\Omega_m$ and $\Omega_\Lambda$ when the mean redshift is
considered:  A universe dominated with $\Omega_\Lambda$ has
a higher number of multiply-imaged sources than a universe
dominated with a large $\Omega_m$. As shown in Fig.~3,
$\bar N$ is essentially a function of the combined 
quantity $\Omega_m-\Omega_\Lambda$, which is mainly
due to a coincidence; the exact dependence is a function of
background source redshifts.
For background source redshifts in the range 1 to 3,
the curves are essentially $\Omega_m-\Omega_\Lambda$.
 We can use this degeneracy in the
lensing rate (e.g., Carroll, Press \& Turner 1992; Kochanek 1993) 
to constrain $\Omega_m-\Omega_\Lambda$ rather than
$\Omega_m$ or $\Omega_\Lambda$ individually.
In Table~3, we list the expected number of strongly lensed radio 
sources along the  $\Omega_m+\Omega_\Lambda=1$ line  as a function of
$\Omega_m-\Omega_\Lambda$. The error bar accounts only for the uncertainty
in redshift. We account for the error in $F$ when cosmological
parameters are estimated, which is facilitated by the
fact that the number of expected lensed sources 
is directly proportional to $F$.

\begin{figure}[ht]
\vspace{6cm}
\caption{Expected number of lensed radio sources, $\bar N$ 
as a function of $\Omega_m$ and $\Omega_\Lambda$.
$\bar N$ is constant along lines of constant $\Omega_m-\Omega_\Lambda$,
allowing for direct constraints on this quantity. Shown here
is the expected number of lensed sources for model A
in Falco et al. 1997}
\end{figure}

\begin{table}
\caption[]{Predicted number of lensed radio sources in the CLASS survey}
\begin{flushleft}
\begin{tabular}{cc}
\noalign{\smallskip}
\hline
\noalign{\smallskip}
$\Omega_m-\Omega_\Lambda$ & $\bar N$ \\                                        
\noalign{\smallskip}
\hline
\noalign{\smallskip}
-1.0 & $55.1^{+8.0}_{-22.5}$\\
-0.8&  $30.7^{+2.4}_{-10.8}$\\
-0.6& $21.5^{+1.3}_{-7.0}$\\
-0.4& $16.6^{+0.8}_{-5.2}$\\
-0.2& $13.5^{+0.5}_{-4.1}$\\
0.0&  $11.3^{+0.4}_{-3.8}$\\
0.2& $9.7^{+0.3}_{-2.8}$\\
0.4& $8.5^{+0.2}_{-2.4}$\\
0.6 & $7.6^{+0.1}_{-2.2}$\\
0.8 & $6.8^{+0.1}_{-1.9}$\\
1.0&  $6.2^{+0.1}_{-1.8}$\\
\noalign{\smallskip}
\hline
\end{tabular}
\end{flushleft}
\end{table}

\section{Constraints on cosmological parameters}

\subsection{Observed number of lensed radio sources}

As listed in Sect. 1, the CLASS survey has produced six new lensed sources
during the first two sessions of its observations from a sample
of 6500 sources. We assume this number to be our canonical case.
However, to account for the possibility that a subsample of lensed sources
has been missed, possibly due to unaccountable selection effects and
other biases in the lens search strategy, we consider the
possibility that there are total of  ten strongly  lensed sources
within this sample and study the variation in cosmological
parameters from the canonical case. We also study the case when
there are four lensed sources.

\subsection{Constraints on $\Omega_m-\Omega_\Lambda$}

We follow CQM to constrain the quantity $\Omega_m-\Omega_\Lambda$ by 
comparing the observed and predicted number of lensed radio sources.
We adopt a Bayesian approach, and take a uniform prior for
$\Omega_m - \Omega_\Lambda$  between -1 and +1, since
we do not yet have a precise determination of
this quantity, and because  we do not wish to consider cosmologies in
which either $\Omega_m$ or $\Omega_\Lambda$ lie outside the interval [0,1].
For values outside the interval of -1 to +1 in $\Omega_m-\Omega_\Lambda$,
we  assume an a priori likelihood of zero.
We do not constrain $\Omega_m$ or $\Omega_\Lambda$ separately;
thus, no prior is required for these quantities.
Since the prior for $\Omega_m - \Omega_\Lambda$ is uniform,
the posterior probability density is simply proportional to the
likelihood.

As derived in CQM, the likelihood ${\cal L}$ --- a function of $\Omega_m - \Omega_\Lambda$ --- is the probability of the data, given $\Omega_m - 
\Omega_\Lambda$. The likelihood for $n$ observed sources (at redshifts
$z_j$) when $\bar N$ is expected is given by:
\begin{equation}
\langle {\cal L}(n)\rangle =\prod_{j=0}^n \tau(z_j) \times e^{- \bar N} \times 
\left(1 + \sigma_{\cal F}^2 \left[ \frac{\bar N^2}{2}-n\bar N +\frac{n(n-1)}{2}\right]\right) \; .
\end{equation}
Here, $n$ is the observed number of
strongly lensed radio sources, while $\bar N$ is the
expected number of lensed sources for a given cosmology.
We have also taken into account the uncertainty in $F$ by defining 
${\cal F} \equiv F/0.026$ and taking ${\cal F}$ to have a mean of unity
and standard deviation $\sigma_{\cal F}=0.3$ allowing for a 
30\% uncertainty in $F$.
The factor ${\cal F}$ is then an overall correction to the
expected lensing rate, due to a systematic uncertainty in $F$.

In order to constrain $\Omega_m-\Omega_\Lambda$, we also need the 
redshifts $z_j$ of the observed lensed sources. Using Table~1 of 
Jackson et al. (1998b),
we obtained redshifts for 4 sources. For the 2 sources with no known
redshifts, we assume a source redshift of 1.5, consistent with other
four redshifts as well as the predicted redshift distribution for
radio sources in the CLASS sample. Changing these redshifts to
reasonably different values do not change our constraints
on the cosmological parameters greatly. We assume that there
are 4, 6 and 10 lensed sources in the CLASS sample.
This is primarily to study the variation in cosmological
parameters with the observed number of lensed sources.

\begin{table}
\caption[]{95\% confidence lower limits on $\Omega_m-\Omega_\Lambda$}
\begin{flushleft}
\begin{tabular}{cc}         
\noalign{\smallskip}
\hline
\noalign{\smallskip}
Case &  $\Omega_m - \Omega_\Lambda$ \\            
\noalign{\smallskip}
\hline
\noalign{\smallskip}
n=6, $z_{\rm mid} $& -0.29 \\
n=6, $z_{\rm low}$ & -0.58 \\
n=6, $z_{\rm high}$& -0.23 \\
n=10, $z_{\rm mid}$& -0.58 \\
n=10, $z_{\rm low}$& -0.90 \\
n=10, $z_{\rm high}$& -0.52 \\
n=4, $z_{\rm mid}$& -0.08 \\
n=4, $z_{\rm low}$& -0.28 \\
n=4, $z_{\rm high}$& -0.04 \\
\noalign{\smallskip}
\hline
\end{tabular}
\end{flushleft}
\end{table}

In Table~4, we summarize our 95\% confidence lower limits
on $\Omega_m-\Omega_\Lambda$ for the three cases, with
$z_{\rm mid}$, $z_{\rm low}$, and $z_{\rm high}$ representing
the mid, low and high values for the redshift distribution 
of radio sources. The cumulative probabilities
for observing six lensed sources with the three different estimates
of the redshift distribution is shown in Fig.~4.
As shown and tabulated, our lowest 95\% confidence on
$\Omega_m-\Omega_\Lambda$ is -0.58. 
Accordingly, in a flat universe with $\Omega_m+\Omega_\Lambda=1$, 
$\Omega_\Lambda < 0.79$ (95\% C.L.). 
If the true redshift distribution for faint radio sources
is represented by the high end allowed by the flux$-$redshift
parameterization, then
$\Omega_m-\Omega_\Lambda > -0.23$ and $\Omega_\Lambda < 0.61$
(95\% C.L.). If there are ten lensed
sources in the present sample of $\sim$ 6500 sources in
the CLASS survey, assuming that four lensed sources have
been missed by current searches, 
then $\Omega_\Lambda < 0.95$ (95\% C.L.). These results
are consistent with previous lensing statistics 
with a 95\% confidence upper limit
on $\Omega_\Lambda$ of 0.57 (Kochanek 1996a) and
cosmological parameters based on high redshift 
type Ia Supernovae (Riess et al. 1998). 
However, these 
limits, especially when $n=6$, are marginally inconsistent
with recent estimates of the cosmological parameters based on the combined
cosmic microwave background (CMB) power spectrum analysis
and the high redshift supernovae (Lineweaver 1998; however,
see, Tegmark 1998), with the
current best estimate for
$\Omega_m-\Omega_\Lambda$ of $-0.38 \pm 0.18$ (1$\sigma$).  
To be consistent, we need more than the currently known six
lensed sources within this CLASS sample or a lower
redshift distribution for faint radio sources 
than allowed by current studies ($z \leq 1$, see below) if there
are six or less lensed sources.
The case for a low redshift distribution can be rejected based
on the observed distribution of redshifts for known lensed sources.

As presented in Table 4, our lower limit on $\Omega_m-\Omega_\Lambda$
varies widely with the used parameterization of flux$-$redshift
relation for background radio sources. Since the number of sources
at the lower flux density end are quite large for the CLASS sample and that
the redshift distribution at this low flux density end
is not known, it is necessary that the redshift distribution
be accurately determined to obtain firm limits on cosmological parameters. 
This can be carried out in two ways:
observational determination based on spectroscopic observations
of optical counterparts of radio sources with flux densities
less than 150 mJy or theoretical determinations based
on the number counts and other information (e.g.,
Jackson \& Wall 1998).

A comparison of Fig.~3 and Table~3, suggest that an upper limit
to $\Omega_m-\Omega_\Lambda$ can also be calculated,
in addition to the lower limit.
Since the number of lensed sources does not vary greatly 
when $\Omega_m-\Omega_\Lambda >0$,
such an upper limit is  weak compared to constraints from other
cosmological probes. Also, this upper limit from lensing lies
outside the current range of interest in $\Omega_m-\Omega_\Lambda$ of
-1 to +1, and will depend on what one assumes for a prior
likelihood.
Also, flat cosmologies where $\Omega_m-\Omega_\Lambda$ is greater
than +1 are incompatible with current constraints on 
cosmological parameters (e.g., White 1998). Therefore, we do not
attempt to calculate an upper limit to
$\Omega_m-\Omega_\Lambda$, hence a lower limit to $\Omega_\Lambda$
in a flat universe, using current lensing statistics.

Since the CLASS survey is expected to observe a total of
$\sim$ 10000 sources, it is possibile that the current
constraints on cosmological parameters may be improved.
In order to investigate this possibility, we used the full catalog
of $\sim$ 10500 sources publicly available from
the CLASS collaboration, which includes sources
that are planned to be observed, as well as the $\sim$ 6500
sources in the current sample.
The present constraints on $\Omega_m-\Omega_\Lambda$ do not vary
greatly when the true number of lensed sources for the whole
sample remains to be
what is obtained based on the current observed lensing rate.
If for some reason, the additional sample of $\sim$ 3500
sources were to contain twice as more lensed sources than 
currently expected, then cosmological parameter estimates
change to be consistent with a high $\Omega_\Lambda$.
However, since this is not likely to be the case, unless the source
selection process is heavily affected systematics,
 an increase in the source sample is not likely to 
improve the radio lensing constraints on cosmological
parameters greatly.

\subsection{Constraints on $\Omega_x-w$}

Recent studies have suggested the
existence of a scalar field as viable alternatives to the cosmological
constant
(see, e.g., Caldwell, Dave \& Steinhardt 1998)
 with an equation of state of the
form $w=P_x/\rho_x$, where $P_x$ and $\rho_x$ are the pressure and
density of the scalar-field, 
respectively. Then, as the Universe expands
$\rho_x 
\propto a^{-3(1+w)}$, where $a$ is the scale factor with $a=(1+z)^{-1}$.
The vacuum energy or the cosmological constant corresponds to
$w=-1$, while texture or tangled strings correspond to $w=-1/3$.

We modified  our lensing probability calculation
to account for the scalar field, and recalculated the likelihoods
using the same technique as above. We assume
a flat universe, and take $\Omega_m+\Omega_x=1$, where
$\Omega_x$ is the present day normalized energy density of
th scalar-field component. We assume a prior in which $\Omega_x$ is
uniform between 0 and +1. Our new constraints are summarized in
Fig.~5 on the $\Omega_x-w$ plane. The 95\% and 68\% upper limits 
({\it thick solid lines}) from gravitational lensing 
are labeled 95\% GL and 68\% GL respectively.
The 95\% confidence. upper limit on $\Omega_x$ can be written as
$\Omega_x \la (1.2 -0.5 w^2) \pm 0.05$.

In addition to lensing constraints, we also
show the resulting constraints on $\Omega_x$ and $w$ 
based on type Ia 
supernovae at high
redshifts (95\% SN) as recently 
studied by Garnavich et al. (1998; {\it dot-dashed
lines}) and constant  age  lines of the Universe
as a function of $H_0t_0$ ({\it dotted lines}).
For a Hubble constant, $H_0$, of 65 $\pm$ 10 km s$^{-1}$ Mpc$^{-1}$
(Riess et al. 1998) and a minimum age of the Universe,
$t_0$, from globular clusters
of $12.5 \pm 1$ Gyr (Chaboyer et al.
1996), a conservative lower limit on $H_0t_0$ is $\sim 0.8$ ({\it
dashed line}). For an age of the universe of $\sim$ 15 $\pm$ 2 Gyr,
$H_0t_0 \sim$ 0.90 to 1.15.
These three cosmological probes, Type Ia supernovae,
age and lensed sources, qualitatively agree with each other 
when $0.5 \la \Omega_x \la 0.8$ and $w \la -0.4$.
These combined results rule out tangled strings with an equation of
state of the form  $w=-1/3$ as the scalar-field component. 

As studied by Turner \& White (1997), the combined large scale structure,
age and supernovae data are consistent with $w\sim -0.6$ when
$\Omega_m \sim 0.3$ to 0.4. The constraints on $\Omega_m$ primarily
comes from the clustered mass-density in galaxy clusters 
based on observed baryonic mass 
combined with a nucleosynthesis determined baryonic
mass density of the universe,
$\Omega_b = (0.019 \pm 0.001) h^{-2}$ (e.g., Burles \& Tytler 1997),
resulting in $\Omega_m = (0.28 \pm 0.07) h^{-2/3}$ (Evrard 1997). 
In Fig.~5, We have outlined a conservative 
lower limit on the known matter content
considering these and similar measurements. 
Even with known matter considered,
there is still a large region left on the $\Omega_x-w$ plane
to form a flat universe
allowing various possibilities for the physical nature of the scalar-field 
component.
In future, constraints on the equation of state of the
unknown scalar-field component is
likely to be improved based on CMB anisotropy and
large scale structure data combined with standard cosmological
probes such as high redshift supernovae and gravitational lensing statistics
(see, Hu et al. 1998).

\subsection{Constraints on $\langle z \rangle$}

Since redshifts for radio sources with flux densities less
than 150 mJy are not accurately known, we can use the observed
lensing rate with an assumed  cosmological model to
constrain the redshift distribution of background
sources. This approach is very similar to
the one used by Holz et al. (1998) to constrain the redshift distribution
of gamma-ray burst sources detected by BATSE based on lensing
statistics.
For the purpose of this calculation, we vary the
redshifts of the background sources with flux densities less than 150 mJy,
while keeping the rest fixed at the appropriate redshifts as determined
earlier. Since we are only interested in an upper limit to the average
redshift of  faint radio sources, 
we calculate lensing probabilities 
as a function of $\langle z \rangle$, the {\it effective} average redshift,
under the assumption that all sources are at this redshift. This
methodology parameterizes our ignorance of the faint radio source
redshift distribution (see, Holz et al. 1998 for a discussion).
However, this assumption is incorrect given that at a given flux 
density level, radio sources have a rather broad distribution in redshift.
Nevertheless, our approach is to suggest a possible upper
limit to the mean redshift of this distribution, assuming that
the distribution is not shaped such that more sources are at higher
redshifts than suggested by the mean. Since the lensing rate is
highly nonlinear in redshift, a slowly
decreasing distribution with increasing redshift can increase
the lensing rate faster than the assumed average value.
In general, our upper limit 
underestimates the true average redshift.

In Table~5, we tabulate the 68\% confidence upper
limit on the redshift of the  subsample of sources
with flux densities less than 150 mJy and assuming that the
total number of lensed sources in the current CLASS sample is
either 6 or 10. This upper limit on the redshift is again
a function of $\Omega_m-\Omega_\Lambda$. If there are
6 lensed sources in the present CLASS sample, then
 $ \langle z \rangle 
\, <\, 1.4 + (\Omega_m-\Omega_\Lambda) \pm 0.1$ (68\% C.L.).
Since the observed redshift distribution of known lensed sources in
the CLASS sample are between 1.3 and 1.6 (see, Table 1 in Jackson
et al. 1998b), cosmological 
models where $\Omega_\Lambda$ is greater than 0.8
in a flat universe are disfavored.

\begin{figure}[ht]
\vspace{6cm}
\caption{Cumulative probability distribution for $\Omega_m-\Omega_\Lambda$,
if there are six strongly lensed sources assuming that the redshifts
are distributed according to model A of Falco et al. (1998) ({\it solid}),
higher ({\it dashed}), and lower ({\it dotted}) from this mean (see,
Sect. 3). The intercepts with the horizontal line show the
95\% confidence lower limits on $\Omega_m-\Omega_\Lambda$ (see, also,
Table~4)}
\end{figure}

\begin{table}
\caption[]{68\% upper limit on  $f_{\rm 8.4\, GHz} \leq 150\, {\rm mJy}\,$ source redshift distribution as a function of $\Omega_m-\Omega_\Lambda$} 
\begin{flushleft}
\begin{tabular}{ccc}         
\noalign{\smallskip}
\hline
\noalign{\smallskip}
$\Omega_m-\Omega_\Lambda$ & $n=6$ & $n=10$  \\            
\noalign{\smallskip}
\hline
\noalign{\smallskip}
-1.0 & 0.57 & 0.84 \\
-0.4 & 1.05 & 1.39 \\
 0.0  & 1.44 & 1.78 \\
 0.4  & 1.83 & 2.24  \\
 1.0  & 2.37 & 2.88 \\
\noalign{\smallskip}
\hline
\end{tabular}
\end{flushleft}
\end{table}

\begin{figure}[ht]
\vspace{6cm}
\caption{Constraints on the $\Omega_x-w$ plane based on gravitational lensing,
high redshift Type Ia supernovae, and globular cluster ages.
We have assumed a spatially flat universe such that
$\Omega_m+\Omega_x=1$. The dotted lines are the constant $H_0t_0$ values,
with $H_0t_0 \sim 0.8$ (dashed-line) 
for the minimum age of the universe. The solid lines are the upper limits
based on gravitational lensing statistics at the 68\% and 95\% confidence.
the 95\% confidence contours from high redshift type Ia supernovae
are shown as dot-dashed lines. If the universe is flat, we can put an
upper limit on $\Omega_x$ based on the known matter content of the
universe as outlined and discussed in text}
\end{figure}

\section{Systematic errors}
As presented in Table~4, 
a major uncertainty  
comes from the unknown redshift distribution of faint 
radio sources with flux densities less than 150 mJy. Thus,
it is unlikely that the current observational data 
on radio sources would be useful to constrain
the cosmological parameters based on  CLASS results, unless
the redshift distribution for its sources is properly
established. Since the redshift and flux density 
distributions for radio sources
are strongly correlated, it is not necessary that redshifts for all $\sim$
6500 sources be known. However, redshifts for a representative
subsample of CLASS sources are much needed. The determination
of redshifts would involve the process of optical identifications
and spectroscopic observations; a task that would take a considerable amount
of telescope time and man power to complete.
However, 
as a result of this study, where we have shown
the strong variation in cosmological
parameters with a change in redshift distribution, 
it is necessary that the true redshift
distribution be used to constrain the cosmological parameters.
We strongly recommend that a subsample of sources be
defined from the CLASS catalog  and be followed up optically to obtain
redshifts. The determination of redshifts would also aid
in the estimation of a reliable luminosity function for 
faint radio sources,
which can be used to improve the calculation associated
with the magnification bias.

In addition to redshift distribution, the selection function for  lensed radio
source discovery process is badly needed. For the present calculation,
we have assumed that all lensed sources with image separations
between 0.3 and 6 arcsecs are recovered. However, the selection
function is likely to be complicated; it is much easier to
recognize a four-image lensed source than a doubly-imaged
lensed source. An initial study of such biases
as applied to the JVAS survey could be found in King \& Browne
(1996). However, as discussed earlier, since radio lens
statistics are much 'cleaner' than the optical ones, we do not
expect biases and selection effects to make a big difference
in present constraints on the cosmological parameters, unless
the selection function is heavily affected by such effects.
Nevertheless, King \& Browne (1996) found that the JVAS survey
contains at least a factor of 2 less doubly imaged sources
than expected from SIS and external shear models. Such a result,
if also confirmed for CLASS survey, suggests
the possibility that a large number of doubly-imaged sources
has been missed by the search process or that a bias not yet understood
affects the observations.

Apart from background sources, an understanding of the foreground lenses
is clearly needed. Use of different $F$ values by neglecting
certain galaxy populations can lead to rather diverse results,
as was recently demonstrated by Chiba \& Yoshi (1998).
Current
deep and wide-area redshift surveys are likely to
increase our knowledge on the foreground lenses and their
composition over the next few years. 
A consistent description
of foreground lenses are also needed for the comparative purposes
between optical and radio lensed source sample studies. 
Currently, however, this task is complicated
by individual studies suggesting different values for
the $F$ parameter, resulting from
different luminosity functions used to describe foreground lenses.
With increasing knowledge on the foreground lenses, an ideal approach would be
to use the luminosity function determined from known  
lensing galaxies (e.g., Kochanek et al. 1998), including the
variation of their number density with redshift.

\section{Summary and conclusions}

Using flux density distribution for a sample of $\sim$ 6500 radio sources
observed during the first two sessions of the Cosmic Lens
All Sky Survey, and the observed number of
strongly lensed sources within this sample, 
we have constrained the cosmological parameters.
We have considered the possibility that there are total of
six (as observed), four or ten lensed
sources within this sample to investigate the changes in our
constraints on the cosmological parameters.

We find that the expected number of lensed sources is primarily
a function of $\Omega_m-\Omega_\Lambda$. A comparison of the
predicted number of lensed sources with
the observed number allows us to constrain 
its current value.
Based on the current determinations of the
redshift distribution for radio sources, our 95\% confidence lower
limit on $\Omega_m-\Omega_\Lambda$ is between -0.58 to -0.23
depending on the assumed model for the redshift distribution of
faint radio sources. For a flat universe with $\Omega_m+\Omega_\Lambda=1$, 
then, $\Omega_\Lambda < 0.61$  to 0.79 (95\% C.L.). If there
are ten strongly lensed sources, the 95\% confidence lower limit
on $\Omega_\Lambda$ is between -0.90 and -0.52.
These limits are in agreement with previous limits based on
gravitational lensing (Kochanek 1996a, 1996b; Falco et al. 1998), and
are not in conflict with estimates based on high redshift supernovae
(viz., $\Omega_m - \Omega_\Lambda\sim -0.5\pm 0.4$ [Riess et al. 1998]).

As has been performed recently in various papers, 
combining $\Omega_m-\Omega_\Lambda$ results from high redshift 
supernovae measurements with
$\Omega_m+\Omega_\Lambda$ results from CMB power spectrum analysis
constrains $\Omega_m$ and $\Omega_\Lambda$ separately, with
much higher accuracy than the individual experiments alone. 
We note that gravitational
lensing constraints on $\Omega_m-\Omega_\Lambda$ should also be
considered in such an analysis, to obtain a consistent picture
on the present cosmological parameters.

In addition to the cosmological constant, we have also studied
the possibility of a scalar field responsible for the
additional energy density of the universe. We have shown our
constraints on the $\Omega_x-w$ plane.
If there are six strongly
lensed sources in the present CLASS sample, 
gravitational lensing statistics allow us to rule
out the region with $\Omega_x \ga (1.2 - 0.5 w^2) \pm 0.05$ (95\% C.L.).
We study the region allowed using combined constraints from
gravitational lensing statistics, 
high redshift Type Ia supernovae distances and limits on
the age of the universe from globular clusters.
We have also considered the current estimates on the known
matter content of the universe. 

Since we have assumed a redshift distribution for
radio sources to constrain cosmological models, alternatively, 
we can constrain the redshift  
distribution of faint radio sources based on the observed
gravitational lensing rate and an assumed cosmological model. 
If there are six strongly lensed sources, the 68\% confidence
upper limit on the redshift distribution of flat-spectrum
radio sources with flux
densities less than 150 mJy at 8.4 GHz is
$ \langle z \rangle \, <\, 1.4 + (\Omega_m-\Omega_\Lambda) \pm 0.1$.

In order to obtain a much tighter estimate on $\Omega_m-\Omega_\Lambda$, 
it is essential that the redshift distribution for radio sources at the 
faint flux density levels be observationally 
determined. Also, estimates of the flat-spectrum radio luminosity function
and the selection function in lensed source
discovery process are needed to complete the story.
We strongly recommend that statistically complete 
optical spectroscopic programs be carried out to
obtain redshifts of faint background radio sources.
Until such redshifts are obtained, it is unlikely that a major
improvement could be made with respect to radio lensing constraints
on cosmological parameters.

\begin{acknowledgements}

I thank members of the CLASS collaboration for making their source
catalog publicly available, Phillip Helbig for useful correspondences,
and Jean Quashnock, Cole Miller, Andr\'e Fletcher, and John Carlstrom
for useful discussions. I would also like to thank the anonymous
referee for  constructive comments on the paper.
This work was supported in part by the 
McCormick Fellowship at the University of Chicago,
and a Grant-In-Aid of Research from the National Academy of 
Sciences, awarded through Sigma Xi, the Scientific Research Society.

\end{acknowledgements}

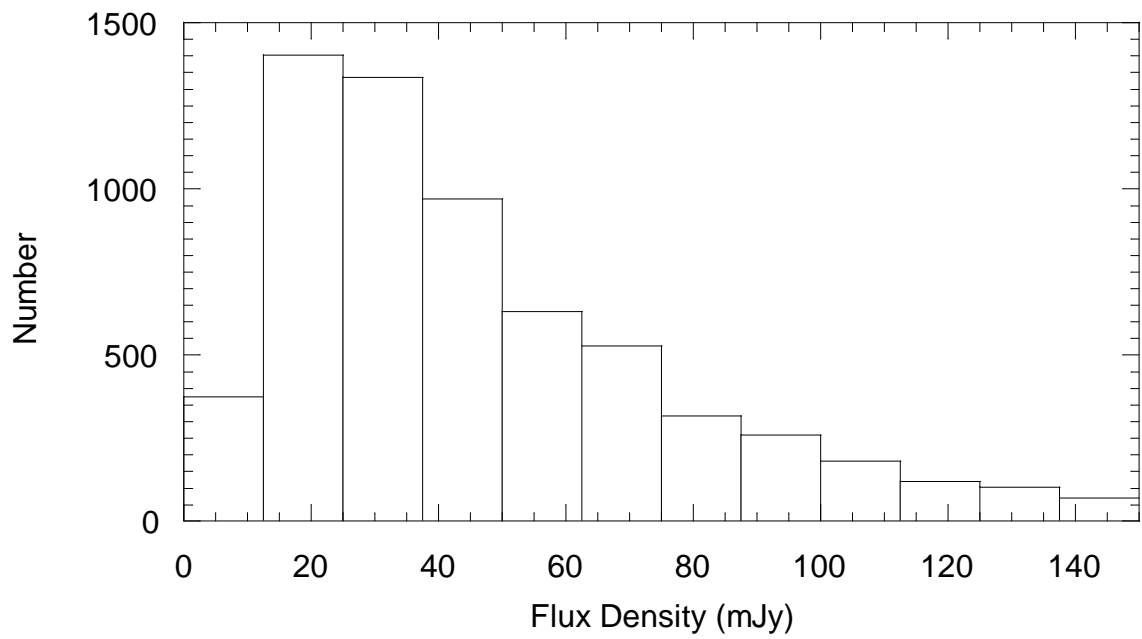
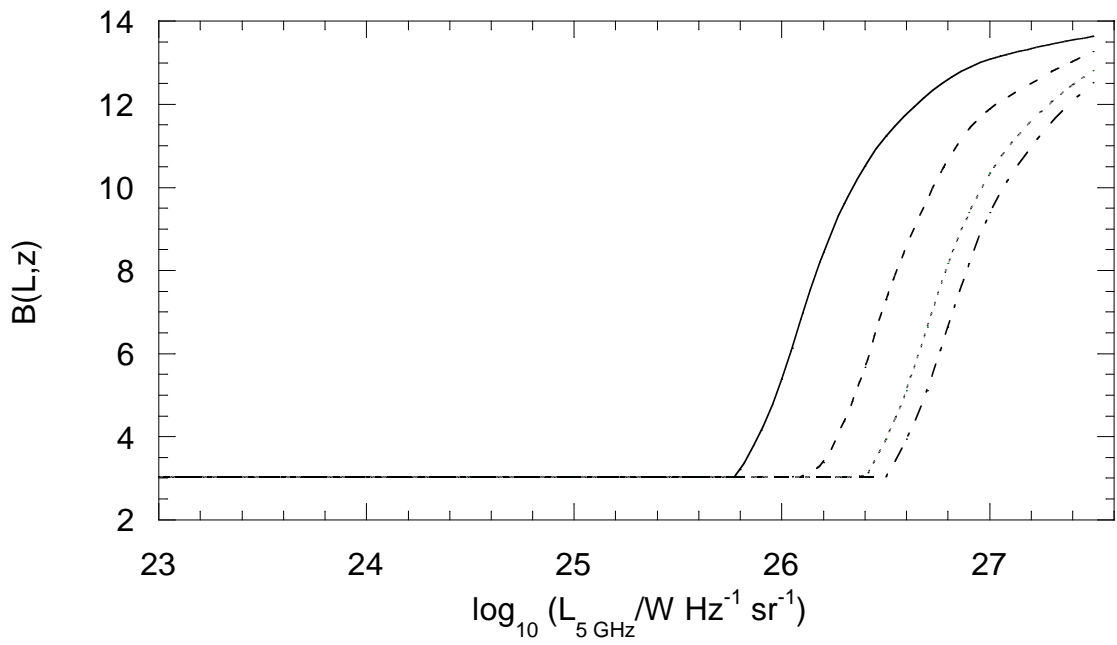
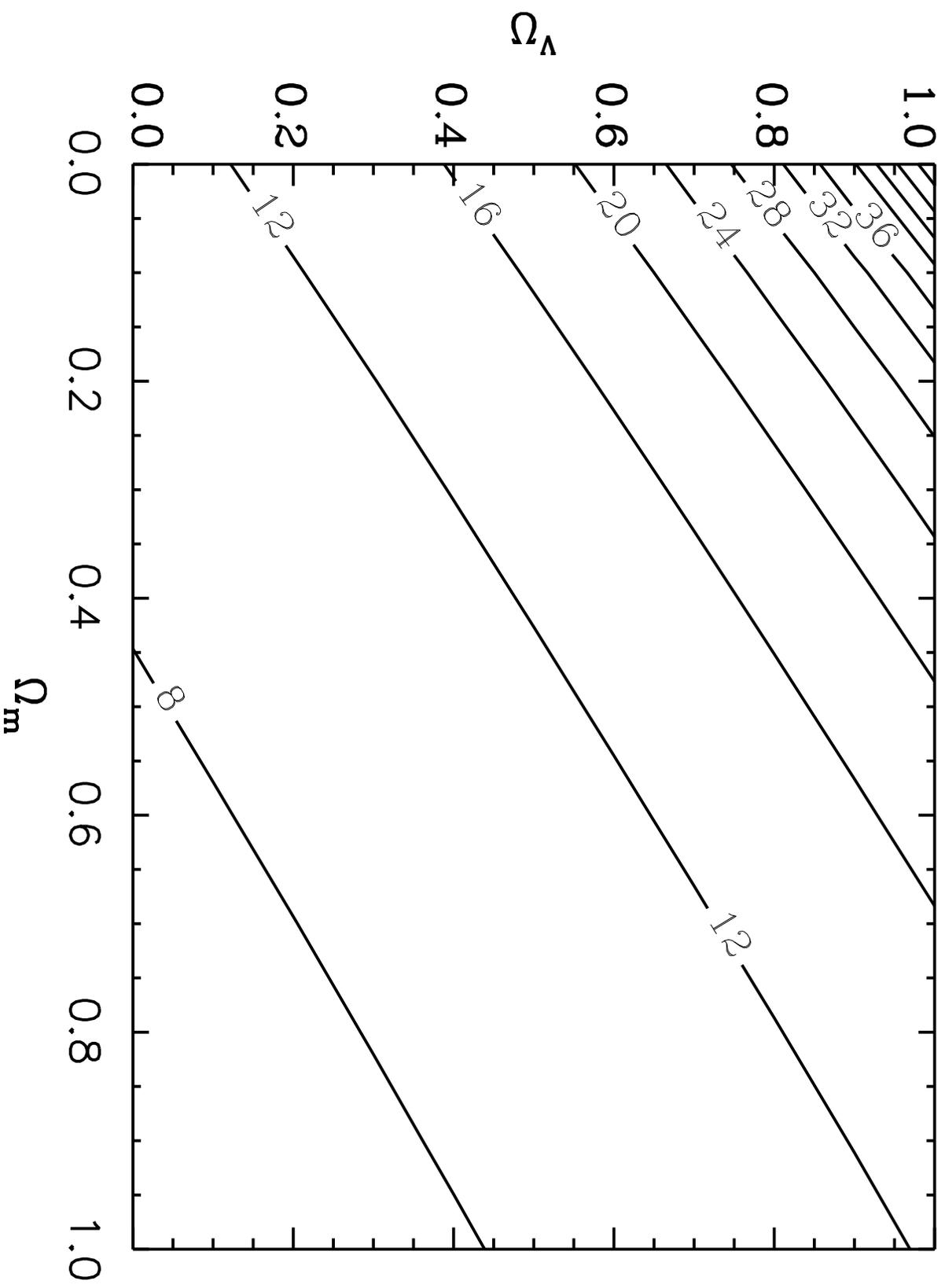
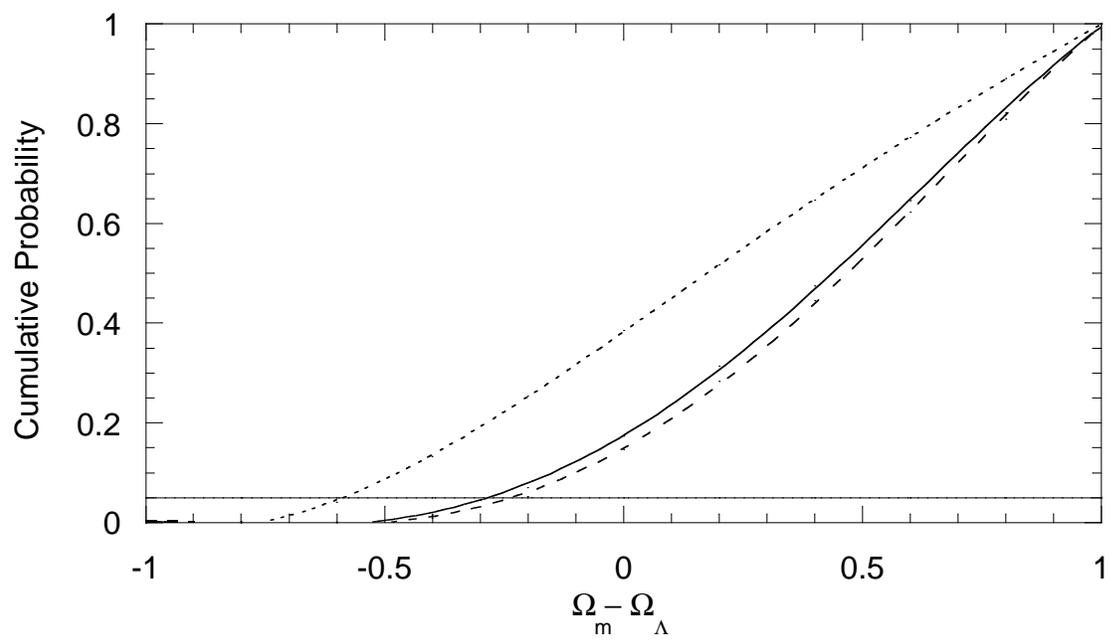
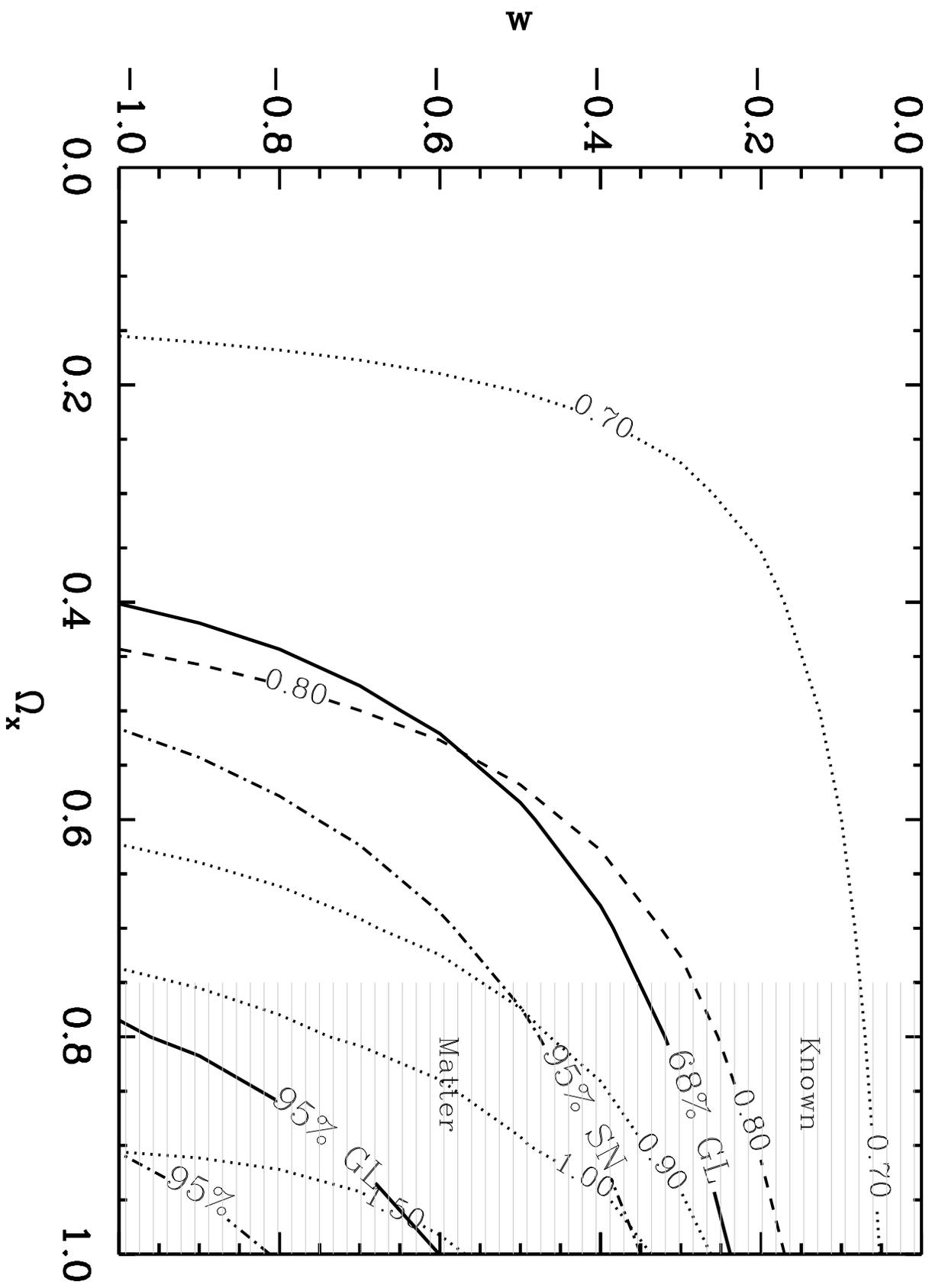
\end{document}